\documentclass{aa}  

\usepackage{graphicx}
\usepackage{txfonts}
\usepackage{natbib,twoopt}
 \usepackage[breaklinks=true]{hyperref} 
 \bibpunct{(}{)}{;}{a}{}{,}    
 \newcommandtwoopt{\citeads}[3][][]{\href{http://adsabs.harvard.edu/abs/#3}%
                                        {\citealp[#1][#2]{#3}}}
 \newcommandtwoopt{\citepads}[3][][]{\href{http://adsabs.harvard.edu/abs/#3}%
                                        {\citep[#1][#2]{#3}}}
 \newcommandtwoopt{\citetads}[3][][]{\href{http://adsabs.harvard.edu/abs/#3}%
                                        {\citet[#1][#2]{#3}}}
 \newcommandtwoopt{\citeyearads}[3][][]%
   {\href{http://adsabs.harvard.edu/abs/#3}{\citeyear[#1][#2]{#3}}}

\begin{document}

   \title{Linear polarization of rapidly rotating ultracool dwarfs}

 \author{P. A. Miles-P\'aez
          \inst{1}\fnmsep\inst{2}
          \and
          M. R. Zapatero Osorio\inst{3}
          \and
          E. Pall\'e\inst{1}\fnmsep\inst{2}
          \and
          K. Pe\~na Ram\'\i rez\inst{1}\fnmsep\inst{2}
          }

   \institute{Instituto de Astrof\'isica de Canarias, La Laguna, E38205 Spain\\
              \email{pamp@iac.es; epalle@iac.es; karla@iac.es}
         \and
             Departamento de Astrof\'isica, Universidad de La Laguna, Av., Astrof\'isico Francisco S\'anchez, s$/$n E38206--La Laguna, Spain\
          \and
             Centro de Astrobiolog\'ia (CSIC-INTA), Carretera de Ajalvir km 4, E-28850 Torrej\'on de Ardoz, Madrid, Spain\\
             \email{mosorio@cab.inta-csic.es}
             }

  \date{Received 2013; accepted 2013}

 
  \abstract
    {}
   {We aim at studying the near infrared linear polarization signal of rapidly rotating ultracool dwarfs with spectral types ranging from M7 through T2 and projected rotational velocities $v$\,sin\,$i\gtrsim30$\,km\,s$^{-1}$. All these dwarfs are believed to have dusty atmospheres and oblate shapes, an appropriate scenario to produce measurable linear polarization of the continuum light.}
   {Linear polarimetric images were collected in the $J$-band for a sample of 18 fast-rotating ultracool dwarfs, five of which were also observed in the $Z$ band using the LIRIS spectrograph on the Cassegrain focus of the 4.2-m William Herschel Telescope. The measured median uncertainty in the linear polarization degree is $\pm$0.13\,\%~for our sample, which allowed us to detect polarization signatures above $\sim$0.39\,\%~with a confidence of $\ge$\,3\,$\sigma$. }
   {About 40\,$\pm$\,15\,\%~of the sample is linearly polarized in the $Z$- and $J$-bands. All positive detections have linear polarization degrees ranging from 0.4\,\%~to 0.8\,\%~in both filters independently of spectral type and spectroscopic rotational velocity. However, simple statistics point at the fastest rotators ($v$\,sin\,$i\gtrsim60$ km\,s$^{-1}$) having a larger fraction of positive detections and a larger averaged linear polarization degree than the moderately rotating dwarfs ($v$\,sin\,$i$\,=\,30--60 km\,s$^{-1}$). Our data suggest little linear polarimetric variability on short time scales (i.e., observations separated by a few ten rotation periods), and significant variability on long time scales (i.e., hundred to thousand rotation cycles), supporting the presence of ``long-term weather'' in ultracool dwarf atmospheres.}
   {}


   \maketitle

\section{Introduction}
Dwarfs with spectral types cooler than M7 (T$_{\rm eff}\le2700$ K; typically referred to as ultracool dwarfs) are believed to undergo the formation of a wide range of atmospheric condensate species (solid and liquid particles) such as corundum (Al$_{2}$O$_{3}$), iron (Fe), enstatite (MgSiO$_{3}$), forsterite (Mg$_{2}$SiO$_{4}$), titanium dioxide (TiO$_{2}$), and gehlenite (Ca$_{2}$Al$_{2}$SiO$_{7}$) among others (\citeads{1997ApJ...480L..39J};\citeads{2001ApJ...556..872A};\citeads{2008MNRAS.391.1854H,2011A&A...529A..44W}). According to models, these condensates may build up into  structures (e.g., ``clouds''), which are located in the outer layers of the atmosphere for T$_{\rm eff}\ge$1300 K and in layers below the visible photosphere for T$_{\rm eff}\lesssim$1300 K \citepads{2001ApJ...556..357A}. Condensates or ``dusty particles''  represent one relevant and yet poorly understood source of opacity in ultracool dwarfs. The presence of atmospheric dust may also polarize the object's output light at particular wavelengths through scattering processes as suggested by the theoretical work of \citepads{2001ApJ...561L.123S}, and observationally demonstrated by the positive detection of linear polarization at optical and near-infrared frequencies (\citeads{2002A&A...396L..35M,2005ApJ...621..445Z,2009A&A...502..929G,2009A&A...508.1423T}; and \citeads{2011ApJ...740....4Z}). Polarization may become a useful tool to comprehend the complexity of ultracool atmospheres.

Additionaly, ultracool dwarfs have high values of projected rotational velocities ($v\,$sin$\,i$) indicating that they are indeed rapid rotators \citepads{2010ApJ...723..684B,2012ApJ...750...79K,2003ApJ...583..451M,2008ApJ...684.1390R,2010ApJ...710..924R,2006ApJ...647.1405Z}. This fact combined with convection and the presence of dust could give rise to intricated atmospheric dynamics, likely generating periodic and non-periodic photometric variability as seen in some late-M, L, and T dwarfs \citepads{2001A&A...367..218B,2001ApJ...557..822M,2004MNRAS.354..378K,2012ApJ...760L..31B,2013AJ....145...71K}. From a theoretical perspective, \citetads{2001ApJ...561L.123S,2003ApJ...585L.155S,2005ApJ...625..996S,2010ApJ...722L.142S,2011Prama..77..157S} and \citetads{2011ApJ...741...59D} predicted that ultracool dwarfs with atmospheric condensates and high $v$\,sin\,$i$'s show measurable linear polarization degrees of typically $\lesssim1\%$ in the optical and near-infrared. Fast rotation induces photospheres into the form of an oblate ellipsoid, and this lack of symmetry leads to incomplete cancellation of the polarization from different areas of the dwarfs surfaces. Gravity is an additional ingredient to take into account since rotationally induced non-sphericity is more favored at lower atmospheric gravities. For a similar amount of dust particles in the ultracool atmospheres it is expected that the largest rotations and lowest gravities produce the largest polarization degrees.

Here, we aim at studying the capabilities of linear polarization in the near infrared to probe the presence of atmospheric condensates in ultracool dwarfs and to shed new light on the dependence of the linear polarization with rotation. We selected a sample of 18 rapidly rotating ultracool dwarfs ($v\,$sin$\,i\gtrsim30$ km\,s$^{-1}$) with spectral types in the interval M7--T2. We report on linear polarimetric imaging observations carried out in the $Z$- and $J$-bands. In Section~\ref{targets} we provide a description of the targets. Section~\ref{observations} presents the observations and data reduction. Polarimetric analysis and discussion are introduced in Sections~\ref{polarimetry} and~\ref{discussion}. Finally, our conclusions are given in Section~\ref{conclusions}.

\section{Target selection \label{targets}}
We selected 18 bright (typically $J < 14.1$ mag) ultracool dwarfs with spectral types from M7 through T2 and published projected rotational velocities $v$\,sin\,$i$ $\ge$ 30 km\,s$^{-1}$ \citepads{2003ApJ...583..451M,2006ApJ...647.1405Z,2008ApJ...684.1390R,2010ApJ...723..684B,2010ApJ...710..924R,2012ApJ...750...79K,2012AJ....144...99D}. All dwarfs are observable from northern astronomical observatories and sufficiently bright at near-infrared wavelengths to achieve accurate polarimetric photometry ($\sigma_P \le 0.2\%$) using short exposures and 4--m class telescopes. They represent $\sim37\%$ of all dwarfs cooler than M7 that have $v$\,sin\,$i$ $\ge$ 30 km\,s$^{-1}$ available in the literature. In Table~\ref{Table1} we provide the targets complete names (abridged names will be used in what follows), their spectral types, near-infrared 2MASS $J$ \citepads{2006AJ....131.1163S} and mid-infrared  {\sl WISE} $W2$ magnitudes, spectroscopic rotational velocities ($v$\,sin\,$i$), and published rotation periods when available. We found {\sl WISE} data for six of our targets (BRI 0021$-$0214, J0036$+$1821, J0700$+$3157, J1254$-$0122, J1835$+$3259, and J2036$+$1051) in \citetads{2012ApJS..201...19D}; we extracted the $W2$ photometry of the remaining sources from the {\sl WISE} catalog \citepads{2010AJ....140.1868W}. For those sources with more than one $v$\,sin\,$i$ measurement in the literature, we provide their weighted mean rotational velocity. Trigonometric and spectrophotometric distance estimates for all objects in our sample are $\leq30$ pc \citepads{2008AJ....136.1290R,2009ApJ...705.1416R,2009AJ....137....1F}. 

\begin{figure}
    \includegraphics[width=0.49\textwidth]{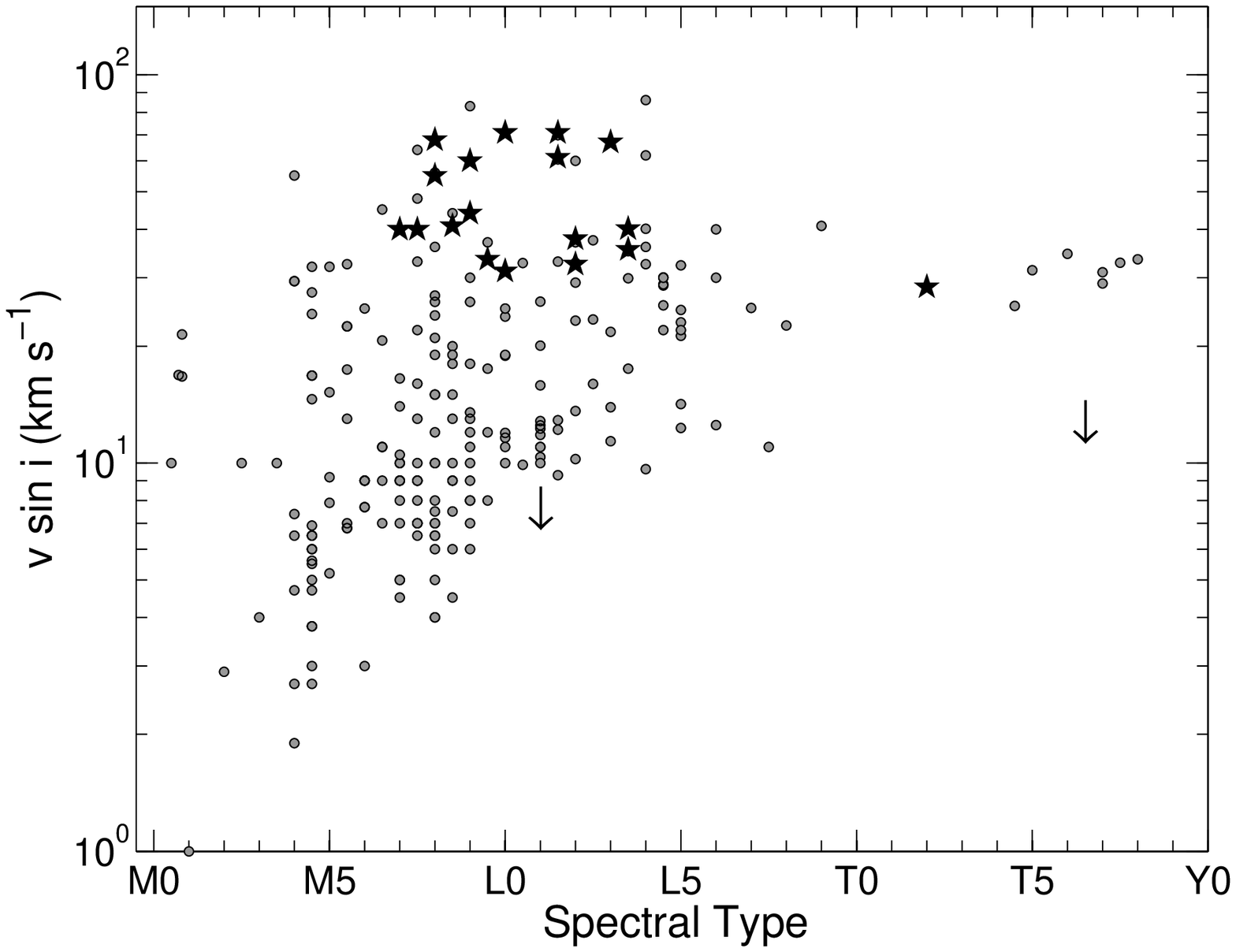}
     \caption{Spectroscopic rotational velocities of M, L and T dwarfs as a function of spectral type. Data taken from \citetads{2010ApJ...723..684B,2012ApJ...750...79K,2003ApJ...583..451M,2008ApJ...684.1390R,2010ApJ...710..924R,2006ApJ...647.1405Z}. Ultracool dwarfs rotate faster with decreasing temperature (and very likely mass). The 18 targets (star symbols) have $v$\,sin$i$\,$\ge$\,30 km\,s$^{-1}$ and spectral types $\ge$\,M7. Typical uncertainties in projected rotational velocities (1--3 km\,s$^{-1}$) and upper limits on the rotational velocities of dwarfs warmer than M7 are not plotted for the clarity of the figure. The great majority of the M0--M4 dwarfs rotate at speeds below $\sim$3 km\,s$^{-1}$.}
              \label{Fig1}%
    \end{figure}

All targets are field dwarfs (e.g., they likely have metal abundance close to solar), and none is reported to be exceptionally young, except for J0045$+$1634, which has an optical spectrum displaying signposts of low-gravity features and likely age and mass around 50--400 Myr and 0.025--0.055 M$_\odot$ (\citeads{2009AJ....137.3345C}; Zapatero Osorio et al$.$ 2013, submitted). The remaining M7--L3.5 and T2 dwarfs have masses estimated at 0.05--0.09 and  0.035--0.065 M$_\odot$, respectively, for an age interval of 0.5--5 Gyr typical of the field \citepads{2003A&A...402..701B}. All dwarfs are either very low-mass stars or brown dwarfs. The great majority of the sample objects (M7--L3.5) have masses around the substellar borderline and quite similar high surface gravities in the interval log\,$g$\,=\,5.1--5.3 (cm\,s$^{-2}$, given by the evolutionary models of \citeads{2003A&A...402..701B}). Figure~\ref{Fig1} illustrates the location of our targets in the projected rotational velocity versus spectral type diagram. As seen from the Figure, our sample includes some of the fastest ultracool rotators known to date. 

In the target sample there are three binary dwarfs, two of which (LP\,349$-$25AB and LP\,415$-$20AB) are resolved and $v$\,sin\,$i$ values are available for each component separately \citepads{2012ApJ...750...79K}. The components of the pairs have similar spectral types  (Table\,\ref{Table1}) and magnitudes: $\Delta$$J=0.84\pm0.15$ mag for LP\,415--20 \citepads{2003ApJ...598.1265S}, and $\Delta$$J=0.35\pm0.03$ mag for LP\,349--25 \citepads{2010ApJ...721.1725D}. The third binary (J0700$+$3157AB) in the target list has differing spectral types and $v$\,sin\,$i$ measured from the combined light at optical and near infrared wavelengths \citepads{2008ApJ...684.1390R,2010ApJ...723..684B}, which is dominated by the primary member since the near-infrared contrast between both components is $\Delta$$J=1.2$ mag \citepads{2006AJ....132..891R} and the contrast is even larger in the visible.

%
\begin{table*}
\caption{List of targets.}
\label{Table1}
\centering
\begin{tabular}{l c c r c c l}
\hline\hline
Name	&SpT	&$J$		&\multicolumn{1}{c}{$W2$} &$v$\,sin\,$i^{\rm a}$		&$T_{\rm rot}$		&Ref. \\
		&		&(mag)	&\multicolumn{1}{c}{(mag)}	&(km\,s$^{-1}$)	&(h)				&	  \\
\hline
2MASS J00192626$+$4614078	&M8		&$12.603\pm0.021$	&$11.001\pm0.020$	&$68\pm10$			& --		& 2	\\
BRI 0021$-$0214			&M9.5		&$11.992\pm0.035$	&$9.900\pm0.019$	&$33.7\pm2.5$			& 4.8		& 2,4,8\\
LP 349$-$25AB			&M8$+$M9	&$10.614\pm0.022$	&$9.054\pm0.020$	&$55\pm2$(A),$83\pm3$(B)	& -- 		& 5 \\	
2MASS J00361617$+$1821104	&L3.5		&$12.466\pm0.027$	&$10.237\pm0.020$	&$35.9\pm2.0$			& 3.1		& 1,3,9\\	
2MASS J00452143$+$1634446	&L2		&$13.059\pm0.022$	&$10.393\pm0.019$	&$32.8\pm0.2$			& -- 		& 1 \\
2MASS J02281101$+$2537380	&L0		&$13.839\pm0.027$	&$11.887\pm0.023$	&$31.2\pm0.8$			& -- 		& 1\\
LP 415$-$20AB			&M7$+$M9.5	&$12.711\pm0.021$	&$11.188\pm0.021$	&$40\pm5$(A),$37\pm4$(B)	& -- 		& 5\\
2MASS J07003664$+$3157266AB	&L3.5$+$L6	&$12.923\pm0.023$	&$10.377\pm0.021$	&$30.1\pm2.0$(A)		& -- 		& 1,3\\
2MASS J08283419$-$1309198	&L2		&$12.800\pm0.030$       &$10.667\pm0.022$	&$30.1\pm2.0$			& 2.9		& 1,3,10 \\
2MASS J11593850$+$0057268	&L0		&$14.084\pm0.028$	&$12.045\pm0.024$	&$71\pm2$			& -- 		& 3\\
2MASS J12545393$-$0122474	&T2		&$14.890\pm0.040$	&$12.396\pm0.025$	&$28.4\pm2.8$			& --		& 7 \\
2MASS J14112131$-$2119503	&M9		&$12.437\pm0.022$	&$10.815\pm0.022$	&$44\pm4$			& --		& 2 \\
2MASS J15010818$+$2250020	&M9		&$11.866\pm0.022$	&$10.053\pm0.020$	&$60\pm2$			& 2.0		& 4,11 \\
2MASS J15210103$+$5053230	&M7.5		&$12.014\pm0.024$	&$10.419\pm0.020$	&$40\pm4$			& --		& 2 \\
2MASS J18071593$+$5015316	&L1.5		&$12.934\pm0.024$	&$10.965\pm0.021$	&$73.6\pm2.2$			& --		& 1,3 \\
2MASS J18353790$+$3259545	&M8.5		&$10.270\pm0.022$	&$8.539\pm0.019$	&$41.2\pm4.7$			& 2.8		& 2,6,12\\
2MASS J20360316$+$1051295	&L3		&$13.950\pm0.026$	&$11.586\pm0.023$	&$67.1\pm1.5$			& --		& 1 \\
2MASS J20575409$-$0252302	&L1.5		&$13.121\pm0.024$	&$10.981\pm0.020$	&$60.6\pm2.0$			& --		& 1,3 \\
\hline\hline
\end{tabular}
\tablefoot{
\tablefoottext{a}{Weighted mean rotational velocity for those targets with more than one measurement. Error bars correspond to the average of individual uncertainties quoted in the literature.}
}
\tablebib{(1) \citetads{2010ApJ...723..684B}; (2) \citetads{2010ApJ...710..924R}; (3) \citetads{2008ApJ...684.1390R}; (4) \citetads{2003ApJ...583..451M}; (5) \citetads{2012ApJ...750...79K}; (6) \citetads{2012AJ....144...99D}; (7) \citetads{2006ApJ...647.1405Z}; (8)  \citetads{2001ApJ...557..822M}; (9) \citetads{2008ApJ...684..644H}; (10) \citetads{2004MNRAS.354..378K}; (11) \citetads{2007ApJ...663L..25H}; (12) \citetads{2008ApJ...684..644H}.
}
\end{table*}
\section{Observations and data reduction \label{observations}}
For the 18 targets, we conducted polarimetric imaging photometry using the $Z$- and $J$-band filters and the Long-slit Intermediate Resolution Infrared Spectrograph (LIRIS; \citeads{2004SPIE.5492.1094M}) attached to the Cassegrain focus of the 4.2-m William Herschel Telescope (WHT) on Roque de los Muchachos Observatory (La Palma, Spain). LIRIS has a 1024\,$\times$\,1024 pixel Hawaii detector covering the spectral range 0.8--2.5 $\mu$m. The pixel projection on the sky is 0\farcs25 yielding a field of view of $4\arcmin.27\times4\arcmin.27$. In its polarimetric imaging mode, LIRIS uses a Wedged double Wollaston device \citepads{1997A&AS..123..589O}, consisting in a combination of two Wollaston prisms that deliver four simultaneous images of the polarized flux at vector angles 0$^{\circ}$ and 90$^{\circ}$, 45$^{\circ}$ and 135$^{\circ}$. An aperture mask 4\arcmin$\,\times\,$1\arcmin~in size is in the light path to prevent overlapping effects between the different polarization vector images. LIRIS does not have an adaptive optics system, binary objects are thus not resolved in our linear polarimetric data. The central wavelengths and widths of the LIRIS $Z$- and $J$-band filters are 1.035$/$0.073 $\mu$m and 1.25$/$0.16 $\mu$m, respectively. 

Observations were carried out during four observing campaings in 2011 December, 2012 June and October, and 2013 January. The journal of the observations is provided in Table~\ref{Table2}. Weather conditions were mainly clear during all campaings and the raw seeing varied between 1\arcsec~and 3\arcsec~in 2011 December and 2012 June and was roughly constant at around 1\arcsec~during the nights of 2012 October and 2013 January.

For each target we obtained linear polarimetric images following a nine--point dither pattern for a proper sky background contribution removal. Typical dither offsets were $\sim$20\arcsec~and $\sim$10\arcsec~along the horizontal and vertical axis. We systematically located our targets on the same spot of the detector, which is close to the center of the LIRIS field of view and optical axis. An example of a LIRIS polarimetric frame obtained through the two Wollaston prisms is given in Figure 1 of \citetads{2011AJ....142...33A}: per frame there are four images of the main source corresponding to vector angles of $0_{TR}^{\circ}$, 90$_{TR}^{\circ}$, 135$_{TR}^{\circ}$, and 45$_{TR}^{\circ}$ from top to bottom, where the subindex $TR$ stands for the telescope rotator angle. The early observations of 2011 December, 2012 June and October were acquired at two different positions of the WHT telescope rotator: $TR$\,=\,0$^{\circ}$ and 90$^{\circ}$. The benefits of this observing strategy are twofold: the flat-fielding effects of the detector are minimized, and the signal-to-noise ratio (S/N) of the polarimetric measurements is improved \citepads{2011AJ....142...33A}. The overheads introduced by the rotation and de-rotation of the telescope rotator were typically about 5 min per target. For the most recent observing run in 2013 January we used two retarder plates, an implementation added to one of the filter wheels of LIRIS during late 2012. These two retarder plates minimize the overheads (by not having to move the telescope rotator) and provide polarimetric images with exchanged orthogonal vector angles, i.e., $0^{\circ}_{90}$, $90^{\circ}_{90}$, $135^{\circ}_{0}$, $45^{\circ}_{0}$, and $0^{\circ}_{0}$, $90^{\circ}_{0}$, $135^{\circ}_{90}$, $45^{\circ}_{90}$ after the aforementioned nomenclature. Typical exposure times per dither ranged from 6 s to 180 s depending on the target brightness, seeing conditions and filter. Total on-source integrations are listed in Table~\ref{Table2} along with the Universal Time (UT) observing dates, filters, air masses, and raw seeing as measured from the averaged full-width-at-half maximum (FWHM) over the reduced images.

Raw data frames were divided into four slices corresponding to the polarimetric vectors, and each slice was reduced following standard procedures for the near infrared using packages within the Image Reduction and Analysis Facility software (IRAF\footnote[1]{IRAF is distributed by the National Optical Astronomy Observatories, which are operated by the Association of Universities for Research in Astronomy, Inc., under cooperative agreement with the National Science Foundation.}). The data reduction steps applied were:
\begin{itemize}
 \item The nine dither frames were median combined to create the sky frame, which was later subtracted from the individual data.
 \item Skyflats were obtained through the polarimetric optics during the sunsets of all observing runs pointing to the east and at high airmasses (sec\,$z \sim 3.0$) to avoid the strong polarization of the sunlight close to the zenith during dusk. All images were divided with the corresponding skyflats normalized to unity to remove detector flat-field variations.
 \item Sky-subtracted and flat-fielded images were registered for a proper alignement.
 \item All aligned images were stacked together to produce deep data.
\end{itemize}

The S/N ratio of the final science images was computed as the ratio between the peak of the flux provided by a Moffat fit to the sources radial profile and the standard deviation of the backgound in a ring of inner radius four times (4\,$\times$) the FWHM and a width of 1\,$\times$\,FWHM. The S/N measurements listed in Table~\ref{Table2} stand for the average values of all four vectors, telescope rotator angles and retarder plates. We note that S/N is always larger for the $J$-band than for the $Z$-band because our targets are very red sources. Nevertheless, the S/N values are in general quite high for the two filters (typically $\ge$500), thus securing the quality of the subsequent photometry. 

Together with the science targets, we also observed polarized and zero-polarized standard stars from the catalogs of \citetads{1992AJ....104.1563S}, \citetads{1992ApJ...386..562W} and non-magnetic white dwarfs, which are supposed to be intrinsically unpolarized. The journal of the observations of the standard stars is provided in Table~\ref{Table2}. These data were reduced in the same manner as the science targets. We used them to control the linear polarization introduced by the telescope and the LIRIS instrument, and to check the efficiency of the LIRIS polarimetric optics (see next Section).

\section{Polarimetric analysis \label{polarimetry}}
\begin{figure}
    \includegraphics[width=0.49\textwidth]{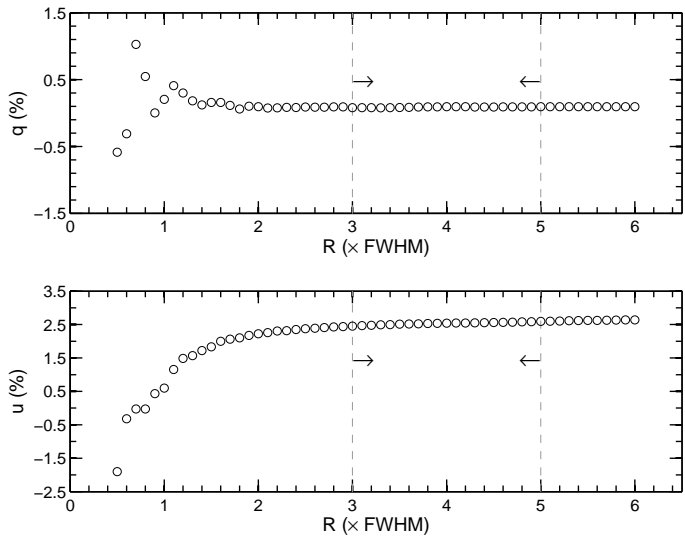}
     \caption{Normalized Stokes Parameters $q$ ({\sl top}) and $u$ ({\sl bottom}) as a function of the aperture radius (in FWHM unit) for the polarized standard star HD\,283855. For this plot, the sky annulus was fixed at 6\,$\times$\,FWHM and 1.5\,$\times$\,FWHM in size. The vertical dashed lines indicate the selected apertures (3--5\,$\times$\,FWHM) to compute the mean Stokes parametes.}
              \label{Fig2}%
    \end{figure}

The normalized Stokes parameters $q$ and $u$ were computed using the flux-ratio method and the equations given in \citetads{2011ApJ...740....4Z}. Fluxes of all polarimetric vectors were measured using the IRAF PHOT package and defining circular photometric apertures of different sizes (from 0.5\,$\times$ to 6\,$\times$\,FWHM with steps of 0.1\,$\times$\,FWHM) and 18 sky rings or annulus of inner radius of 3.5\,$\times$ through 6\,$\times$\,FWHM (steps of 0.5\,$\times$\,FWHM) and widths of 1\,$\times$, 1.5\,$\times$, and 2\,$\times$\,FWHM. The sky annuli account for possible background residuals remaining from the previous data reduction steps. Summarizing, for each polarimetric vector a total of 990 fluxes were computed. In Figure~\ref{Fig2}, we illustrate the resulting normalized Stokes parameters of a polarized standard star as a function of the circular photometric aperture (we fixed the sky annulus for the clarity of the Figure). The final $q$ and $u$ values were obtained by selecting the range of apertures providing a flat distribution of the Stokes parameters (as shown in Figure~\ref{Fig2}), and by averaging all the bracketed $q$ and $u$ measures (including all sky annuli). The selected range of photometric apertures for each science target and standard star is listed in Table~\ref{Table2}: typical photometric aperture size goes from 2\,$\times$ to 4 \,$\times$\,FWHM. The uncertainties associated to $q$ and $u$ were determined as the standard deviation of all selected individual measures. Tables~\ref{Table3} and \ref{Table4} provide the final normalized Stokes parameters and their error bars for both science targets and standard stars. The high S/N photometry leads to small uncertainties in $q$ and $u$, typically below $\pm$\,0.1\,\%.

The degree of linear polarization, $P$, and the linear polarization vibration angle, $\Theta\in$ [0,\,$\pi$]\,rad, are calculated from the $q$ and $u$ normalized Stokes parameters using the equations given in \citetads{2011ApJ...740....4Z}. Error bars associated to $P$ are computed as the quadratic sum of the $q$ and $u$ quoted uncertainties plus the uncertainty of 0.10\,\%~introduced by a possible instrumental linear polarization (see below). The error in the polarization vibration angle is obtained from the following expression:
   \begin{equation}
      \sigma_{\Theta} =\sqrt{(28.65\,\sigma_{P}/P)^{2}+\sigma_{\Theta_{0}}^{^2}  }
   \end{equation}
where $\sigma_{\Theta}$ and $\sigma_{\Theta_{0}}$ are in deg, and the factor $28.65\,\sigma_{P}/P$ comes from \citetads{1974apoi.book..361S} and \citetads{1974ApJ...194..249W}. All measured $P$ and $\Theta$ values along with their associated uncertainties are provided in Tables~\ref{Table3} and ~\ref{Table4}. We note that $P$ and $\Theta$ error bars are typically dominated by the instrumental polarization uncertainty. The median error in $P$ is $\pm$0.13\,\%~for both $Z$- and $J$-band measurements. 

\begin{table*}
\caption{Observing log.}
\label{Table2}
\centering
\begin{tabular}{l c c c c c r c}
\hline\hline
Object 	&Aperture$^{\rm a}$	& Obs. date 	&Filter 	&Exposure Time$^{\rm b}$ 	&FWHM	&\multicolumn{1}{c}{S/N}	&Airmass\\
 		&(FWHM)			&(UT)		& 		& (s)			&(")	&	&	\\
\hline
J0019$+$4614	&3--4   	&2012 Oct 06 	&$J$ 	&1$\times$9$\times$30, 1$\times$9$\times$50   	&0.9 	&2500 	&1.40--1.33\\
                       	 	&2--3 	&2012 Oct 07 	&$Z$ 	&1$\times$9$\times$120, 1$\times$9$\times$120	&1.0 	&1750 	&1.06--1.03\\
BRI 0021$-$0214 	&3--4 	&2012 Oct 06	&$J$ 	&1$\times$9$\times$20, 1$\times$9$\times$20		&0.9		&3540	&1.60--1.57\\
LP349$-$25AB		&3--4	&2012 Oct 06	&$J$		&2$\times$9$\times$6, 2$\times$9$\times$6		&0.7		&10900	&1.04--1.02\\
          			&3--4	&2012 Oct 07	&$Z$	&1$\times$9$\times$50, 1$\times$9$\times$50		&0.8		&7740	&1.05--1.03\\
J0036$+$1821	&3--4	&2012 Oct 06	&$J$		&2$\times$9$\times$35, 2$\times$9$\times$35		&0.7		&6580	&1.03--1.02\\
J0045$+$1634	&3--4	&2012 Oct 06	&$J$ 	&1$\times$9$\times$50, 1$\times$9$\times$50		&1.0		&2570	&1.02--1.04\\
J0228$+$2537	&3--4	&2012 Oct 06	&$J$		&1$\times$9$\times$70, 1$\times$9$\times$70   	&0.9		&1100	&1.04--1.02\\
LP415$-$20AB 	&3--4	&2012 Oct 06	&$J$		&1$\times$9$\times$40, 1$\times$9$\times$60		&0.8		&2800	&1.02--1.04\\
                            	&3--4	&2012 Oct 07	&$Z$	&1$\times$9$\times$120, 1$\times$9$\times$120	&0.8		&2000	&1.04--1.09\\
J0700$+$3157AB 	&3--4	&2012 Oct 06	&$J$		&1$\times$9$\times$60, 1$\times$9$\times$60		&0.8		&3070	&1.07--1.04\\
J0828$-$1309		&2--3	&2011 Dec 31 &$J$		&1$\times$9$\times$120, 1$\times$9$\times$80	&0.9		&4420	&1.43--1.56\\
                            	&3--4 	&2013 Jan 29	&$J$		&1$\times$9$\times$60, 1$\times$9$\times$60		&0.8		&1820	&1.34--1.35\\
J1159$+$0057	&2--3	&2012 Jun 16	&$J$		&1$\times$9$\times$120, 1$\times$9$\times$120	&1.5		&380	&1.33--1.71\\
                       		&1--2	&2013 Jan 29	&$J$		&1$\times$9$\times$120, 1$\times$9$\times$120	&1.1		&425	&1.40--1.30\\
J1254$-$0122		&2.5--3.5&2013 Jan 29	&$J$		&2$\times$9$\times$120, 2$\times$9$\times$120	&0.9		&470	&1.44--1.21\\
J1411$-$2119		&3--4	&2012 Jun 16	&$J$		&2$\times$9$\times$120, 2$\times$9$\times$120	&2.5		&460	&1.85--2.79\\
J1501$+$2250	&3--4	&2013 Jan 29	&$J$		&1$\times$9$\times$60, 1$\times$9$\times$60		&0.9		&2930	&1.35--1.27\\
J1521$+$5053	&3--4	&2013 Jan 29	&$J$		&1$\times$9$\times$60, 1$\times$9$\times$60		&0.8		&3950	&1.31--1.26\\
J1807$+$5015	&3--4	&2012 Jun 15	&$J$		&1$\times$9$\times$120, 1$\times$9$\times$120	&1.0		&2710	&1.14--1.21\\
                       		&2.5--3.5&2012 Oct 07	&$Z$	&1$\times$9$\times$120, 1$\times$9$\times$180	&0.8		&1500	&1.26--1.44\\
J1835$+$3259	&2--3	&2012 Jun 15	&$J$		&1$\times$9$\times$5, 1$\times$9$\times$5		&0.8		&6100	&1.12--1.15\\
J2036$+$1051	&2--3	&2012 Jun 15	&$J$		&1$\times$9$\times$90, 1$\times$9$\times$90		&1.0		&870	&1.06--1.08\\
                       		&2--3	&2012 Oct 07	&$Z$	&1$\times$9$\times$180, 1$\times$9$\times$180	&1.1		&516	&1.12--1.28\\
J2057$-$0252		&3--4	&2012 Oct 06	&$J$		&1$\times$9$\times$40, 1$\times$9$\times$30		&0.7		&2617	&1.25--1.21\\
\hline
SA29$-$130$^{\rm c}$	&3--4	&2013 Jan 29	&$J$		&1$\times$9$\times$100, 1$\times$9$\times$100	&0.8		&1060	&1.05--1.04\\
GJ\,3805$^{\rm c}$		&2--3	&2012 Jun 15	&$J$		&1$\times$9$\times$100, 1$\times$5$\times$100	&1.0		&992	&1.14--1.13\\
BD$+$284211$^{\rm c}$&3--4	&2012 Oct 06	&$J$		&1$\times$9$\times$7, 1$\times$9$\times$7		&0.7		&4900	&1.06--1.05\\
               			&3--4	&2012 Oct 07	&$Z$	&1$\times$9$\times$60, 1$\times$9$\times$60		&0.9		&8423	&1.10--1.05\\
Feige 110$^{\rm c}$	&2--3	&2011 Dec 31	&$J$		&1$\times$9$\times$90, 1$\times$5$\times$90		&1.4		&1100	&1.31--1.41\\
              			&2--3	&2012 Jun 16	&$J$		&1$\times$9$\times$120, 1$\times$5$\times$120	&1.7		&1240	&1.40--1.28\\
HD\,283855$^{\rm d}$	&3--5	&2012 Oct 06	&$J$		&2$\times$9$\times$3, 2$\times$9$\times$3		&0.7		&12600	&1.06--1.04\\
            			&3--4	&2012 Oct 07	&$Z$	&2$\times$9$\times$6, 2$\times$9$\times$6		&0.7		&7400	&1.00--1.01\\
HRW\,24$^{\rm d}$		&2--3	&2011 Dec 31	&$J$		&1$\times$9$\times$20, 1$\times$9$\times$20		&0.9		&3414	&2.54--2.92\\
          			&2--3	&2013 Jan 29	&$J$		&1$\times$9$\times$10, 1$\times$9$\times$10		&0.7		&1230	&1.36--1.37\\
\hline\hline
\end{tabular}
\tablefoot{
\tablefoottext{a}{Range of circular photometric apertures (in units of FWHM) used to determine the normalized Stokes parameters.}
\tablefoottext{b}{Number of times the 9-dither patter is repeated $\times$ 9 images $\times$ single exposure time, one column for each telescope rotator angle and retarder plate. With very few exceptions, the exposure times were identical for the two telescope rotator angles and retarder plates.}
\tablefoottext{c}{Zero-polarized standard star.} \tablefoottext{d}{Polarized standard star \citepads{1992ApJ...386..562W}.}
}
\end{table*}
%

Before proceeding to the discussion of the polarimetric measurements, we dealt with the data systematics. Because the polarimetric images were acquired with two slightly different observing configurations (two angles of the telescope rotator versus retarder plates), first of all we did secure that both configurations are compatible and yield the same results. The non-polarized standard star SA29$-$130 was observed using the retarder plates in 2013 January (Table~\ref{Table4}). The measured linear polarization in the $J$-band is $P\,=0.05\pm0.16\,\%$ compatible with null polarization at 1.25 $\mu$m and suggesting that the ``retarding plates'' configuration does not introduce significant instrumental polarization. Observations of other unpolarized standard stars in previous campaings and using the ``two angles of the telescope rotator'' configuration also yield negligible instrumental linear polarization within an uncertainty of $\pm$0.1\,\%, in agreement with previous bibliography \citetads{2011AJ....142...33A,2011ApJ...740....4Z}. In addition, the polarized standard star HRW\,24 was observed with the two instrumental configurations in 2011 December and 2013 January (Table~\ref{Table4}), finding that its $J$-band linear polarization degree and polarization vibration angle are consistent at the 1-$\sigma$ level with each other and with the published data. Therefore, we concluded that the two instrumental configurations do not introduce a relative bias in our measurements, and that all observations can be considered ``homogeneously'' obtained in the following discussion.

We also checked that our previously described data reduction procedure did not introduce an extra polarization signal in the measurements. The raw images of the standard stars were re-reduced skipping the flat-fielding correction step and/or using skyflat images obtained without the polarimetric optics. We observed that for polarized stars the new reduction delivered $P$ and $\Theta$ values far from what is tabulated in the literature (literature measurements are given in Table~\ref{Table4} permitting a proper comparison with our determinations). In addition, zero-polarized stars appeared linearly polarized with degrees of $\sim0.3-0.4\%$ in the newly reduced $J$-band images. The situation dramatically changed when we applied the standard data reduction procedure that includes flat-field correction using skyflat images obtained with the polarimetric optics. From Table~\ref{Table4} and the observations of unpolarized standard stars, we derived that LIRIS on the Cassegrain focus of the WHT has small (if any) instrumental linear polarization, likely below 0.1--0.2\%~at both the $Z$- and $J$-band filters.

The response or efficiency of LIRIS polarimetric optics was controlled by the observations of the strongly polarized standard stars HD\,283855 and HRW\,24. As shown in Table~\ref{Table4}, our $J$-band linear polarization degrees are in good agreement (at the 1-$\sigma$ level) with the data from the literature, indicating no correction factor for efficiency loss at the $J$-band. Unfortunately, there are no published polarimetric indices in the $Z$-filter for the standard stars. However, our $Z$-band measurement for HD\,283855, $P_{Z}=\, 3.45\pm0.1\,\%$ and $\Theta_{Z}\,=\,40.4\,\pm\,0.2$ deg, lies intermediate between the literature linear polarization degrees in the $I$- and $J$-filters, i.e., $P_{Z,{\rm lit}}=\, 3.3\pm0.1\,\%$ and $\Theta_{Z,{\rm lit}}\,=\,46\,\pm\,1$ deg \citetads{1992ApJ...386..562W}, as expected. Therefore, we did not apply any correction factor due to efficiency loss to our $Z$-band measurements. There is, however, a zero-point correction to be applied to the position angle of the polarization vibration in the $J$-band, which is measured at $\Theta_{\rm o}\,=\,+4.4\,\pm\,1.3$ deg, indeed very similar to the quantity $\Theta_{o}=+4.46\pm1.5$ deg given in \citetads{2011ApJ...740....4Z}. For this, we did not take into account the observations of the standard star HD\,283855 since its $q$ values are close to 0.0, thus introducing large uncertainty in the determination of $\Theta_{o}$. There are not sufficient data to determine the zero-point of the polarization vibration angle for the $Z$-band. This correction is wavelength dependent, yet given the proximity in wavelength of the filters $Z$ and $J$, we applied the same $\Theta_{o}$ to $Z$- as to the $J$-band.

We discarded any correlation between our $P$ measurements and the airmass at which the targets were observed. The $P$ versus airmass plane was divided into four quadrants with an origin at the median values of airmass and $P$ (1.17 and 0.38\,\%, respectively). In case of any positive correlation, measurements would be located in preferred quadrants. We did not observe any grouping of our data in any of the quadrants. The Pearson's $r$ correlation coefficient is $r=0.13\pm0.21$ for the $P$ measurements, implying that only $\sim1.7\%$ of the data could be explained by a model of linear correlation with airmass. Therefore, we concluded that the polarization measurements are not biased by the airmass of the observations.

Because the linear polarization degree $P$ is always a positive quantity, small values of $P$ and values of $P$ affected by poor signal-to-noise-ratio data are statistically biased toward an overestimation of the true polarization (see \citeads{1985A&A...142..100S}). We applied the equation given by \citetads{1974ApJ...194..249W} to derive the debiased linear polarization degree, $p^{*}$, by taking into account the measured $P$ and its associated uncertainty:
 \begin{equation}
      p^{*}\,=\,\sqrt{P^{2}-\sigma_{P}^{2}}
   \end{equation}
The debiased linear polarization degrees $p^{*}$ and polarization vibration angles are provided in Tables~\ref{Table3} and ~\ref{Table4} for the science targets and standard stars, respectively. At high values of polarization, changes are negligible, i.e., this correction does not affect the positive detection of linear polarization in eight ultracool dwarfs in the sample.
\begin{table*}
\caption{Linear polarimetry photometry of science targets.}\label{Table3}
\centering
\begin{tabular}{l c c r r c c r} 
\hline\hline             
Object	&Filter 	&Obs. time			&\multicolumn{1}{c}{$q$}	&\multicolumn{1}{c}{$u$}	&$P$	&$p^{*}$	&\multicolumn{1}{c}{$\Theta$}\\
  		&  		&(JD$-$2450000.5) 	&\multicolumn{1}{c}{($\%$)}	&\multicolumn{1}{c}{($\%$)}	&($\%$)	&($\%$)	&\multicolumn{1}{c}{(deg)} \\
\hline
\hline
J0019$+$4614	&$J$		&6206.8719	&$-0.40\pm0.06$		&$0.11\pm0.09$		&0.41$\,\pm\,$0.15	&0.38$\,\pm\,$0.15	&--\\
          			&$Z$	&6207.9954	&$0.31\pm0.07$		&$-0.49\pm0.05$		&0.58$\,\pm\,$0.13	&0.57$\,\pm\,$0.13	&146.8$\,\pm\,$6.7\\
BRI0021$-$0214	&$J$		&6206.8988	&$-0.10\pm0.03$		&$0.10\pm0.03$		&0.14$\,\pm\,$0.11	&0.09$\,\pm\,$0.11	&--\\
LP349$-$25AB		&$J$		&6206.9830	&$0.14\pm0.02$		&$0.15\pm0.05$		&0.21$\,\pm\,$0.11	&0.18$\,\pm\,$0.11	&--\\
          			&$Z$	&6208.4696	&$-0.22\pm0.02$		&$-0.02\pm0.02$		&0.22$\,\pm\,$0.10	&0.20$\,\pm\,$0.10	&--\\
J0036$+$1821	&$J$		&6207.0147	&$-0.11\pm0.04$		&$0.20\pm0.03$		&0.23$\,\pm\,$0.11	&0.20$\,\pm\,$0.11	&--\\
J0045$+$1634	&$J$		&6207.0275	&$-0.01\pm0.03$		&$-0.08\pm0.03$		&0.08$\,\pm\,$0.11	&0.00$\,\pm\,$0.11	&--\\
J0228$+$2537	&$J$		&6207.0697	&$-0.15\pm0.08$		&$-0.35\pm0.06$		&0.38$\,\pm\,$0.14	&0.35$\,\pm\,$0.14	&--\\
LP415$-$20AB		&$J$		&6207.2099	&$0.41\pm0.06$		&$-0.01\pm0.05$		&0.41$\,\pm\,$0.12	&0.39$\,\pm\,$0.12	&174.9$\,\pm\,$9.0\\
          			&$Z$	&6208.2221	&$0.39\pm0.06$		&$0.01\pm0.05$		&0.39$\,\pm\,$0.12	&0.38$\,\pm\,$0.12	&176.3$\,\pm\,$9.4\\
J0700$+$3157AB	&$J$		&6207.2425	&$-0.50\pm0.08$		&$-0.03\pm0.04$		&0.50$\,\pm\,$0.14	&0.48$\,\pm\,$0.14	&87.3$\,\pm\,$7.8\\
J0828$-$1309		&$J$		&5927.1754	&$-0.15\pm0.02$		&$0.07\pm0.02$		&0.17$\,\pm\,$0.10	&0.14$\,\pm\,$0.10	&--\\
            			&$J$		&6321.0538	&$0.26\pm0.10$		&$0.09\pm0.06$		&0.27$\,\pm\,$0.15	&0.22$\,\pm\,$0.15	&--\\
J1159$+$0057	&$J$		&6094.9304	&$0.23\pm0.08$		&$0.50\pm0.08$		&0.55$\,\pm\,$0.15	&0.53$\,\pm\,$0.15	&28.3$\,\pm\,$8.0\\
          			&$J$		&6321.1119	&$0.42\pm0.07$		&$0.05\pm0.05$		&0.42$\,\pm\,$0.13	&0.40$\,\pm\,$0.13	&179.0$\,\pm\,$9.0\\
J1254$-$0122		&$J$		&6321.1575	&$-0.31\pm0.25$		&$-0.01\pm0.21$		&0.31$\,\pm\,$0.34	&0.00$\,\pm\,$0.34	&--\\
J1411$-$2119		&$J$		&6094.9835	&$0.30\pm0.06$		&$-0.41\pm0.13$		&0.51$\,\pm\,$0.17	&0.48$\,\pm\,$0.17	&148.7$\,\pm\,$10.0 \\
J1501$+$2250	&$J$		&6321.1950	&$-0.10\pm0.06$		&$0.51\pm0.04$		&0.52$\,\pm\,$0.12	&0.51$\,\pm\,$0.12	&46.2$\,\pm\,$7.0\\
J1521$+$5053	&$J$		&6321.2133	&$0.49\pm0.07$		&$0.36\pm0.04$		&0.61$\,\pm\,$0.13	&0.60$\,\pm\,$0.13	&13.8$\,\pm\,$6.2\\
J1807$+$5015	&$J$		&6094.1371	&$-0.54\pm0.03$		&$0.41\pm0.04$		&0.68$\,\pm\,$0.11	&0.67$\,\pm\,$0.11	&67.0$\,\pm\,$5.0\\
         			&$Z$	&6207.8720	&$0.07\pm0.09$		&$0.20\pm0.07$		&0.21$\,\pm\,$0.15	&0.15$\,\pm\,$0.15	&--\\
J1835$+$3259	&$J$		&6094.1755	&$-0.02\pm0.02$		&$-0.12\pm0.01$		&0.12$\,\pm\,$0.10	&0.07$\,\pm\,$0.10	&--\\
J2036$+$1051	&$J$		&6094.1880	&$-0.10\pm0.11$		&$-0.22\pm0.05$		&0.24$\,\pm\,$0.16	&0.18$\,\pm\,$0.16	&--\\
         			&$Z$	&6207.9429	&$-0.78\pm0.18$		&$0.07\pm0.20$		&0.78$\,\pm\,$0.28	&0.73$\,\pm\,$0.28	&--\\
J2057$-$0252		&$J$		&6206.8253	&$0.43\pm0.08$		&$-0.05\pm0.08$		&0.43$\,\pm\,$0.15	&0.40$\,\pm\,$0.15	&--\\
\hline
\hline
\end{tabular}
\end{table*}
\begin{table*}
\caption{Linear polarimetry photometry of standard stars.}
\label{Table4}
\centering
\begin{tabular}{l c r r c c c c}
\hline\hline
Object		&Filter	 &\multicolumn{1}{c}{$q$}				&\multicolumn{1}{c}{$u$}		&$P$	&$p^{*}$	&$\Theta$	&$P_{\rm lit},\Theta_{\rm lit}^{\rm a}$\\
			&		 &\multicolumn{1}{c}{($\%$)}				&\multicolumn{1}{c}{($\%$)}		&($\%$)	&($\%$)	&(deg)		&(\%, deg) \\
\hline
SA29$-$130	&$J$		&$-0.03\pm0.12$	&$0.05\pm0.11$	&0.05$\,\pm\,$0.16		&-- 	&-- 	&--\\
GJ\,3805		&$J$		&$-0.04\pm0.26$	&$-0.06\pm0.06$	&0.08$\,\pm\,$0.26		&-- 	&-- 	&--\\
BD$+$284211&$J$	&$0.00\pm0.05$	&$0.11\pm0.04$	&$0.11\pm0.07$	&--      &-- 	&--\\
		&$Z$		&$0.06\pm0.03$	&$-0.09\pm0.05$	&0.12$\,\pm\,$0.06		&-- 	&-- 	&--\\
Feige 110	&$J^{\rm b}$	&$0.08\pm0.13$	&$0.02\pm0.08$	&0.09$\,\pm\,$0.15	&-- 	&-- 	&--\\
		&$J^{\rm c}$	&$0.03\pm0.04$	&$-0.06\pm0.05$	&0.07$\,\pm\,$0.07	&-- 	&-- 	&--\\
HD\,283855	&$J$		&$0.19\pm0.02$	&$2.07\pm0.02$	&2.08$\,\pm\,$0.10		&2.07$\,\pm\,$0.10	&42.4$\,\pm\,$1.5	&2.58$\pm$0.05, 46$\pm$1\\
 		&$Z$		&$0.56\pm0.01$	&$3.41\pm0.03$	&3.46$\,\pm\,$0.10		&3.45$\,\pm\,$0.10	&40.4$\,\pm\,$1.0	&3.3$\pm$0.1, 46$\pm$1$^{\rm d}$\\
HRW\,24		&$J^{\rm b}$	&$-2.19\pm0.03$	&$-0.03\pm0.02$	&2.19$\,\pm\,$0.11		&2.18$\,\pm\,$0.11	&90.4$\,\pm\,$1.5	&2.10$\pm$0.03, 86$\pm$1\\
 		&$J^{\rm e}$	&$-2.34\pm0.09$	&$-0.03\pm0.08$	&2.34$\,\pm\,$0.16		&2.33$\,\pm\,$0.16	&90.4$\,\pm\,$1.9	& \\
\hline\hline
\end{tabular}
\tablefoot{
\tablefoottext{a}{Literature data from \citetads{1992ApJ...386..562W}.}
\tablefoottext{b}{2011 December.}
\tablefoottext{c}{2012 June.}
\tablefoottext{d}{Computed from the literature $I$ and $J$ data for a wavelength of 1.03 $\mu$m (LIRIS $Z$-band).}
\tablefoottext{e}{2013 January.}
}
\end{table*}
%
\section{Discussion \label{discussion}}
We adopted the 3-$\sigma$ criterion to identify positive detection of linear polarization, i.e.,  $P/\sigma_P\,\ge\,3$. This criterion has been extensively used in previous works \citepads{2002A&A...396L..35M, 2005ApJ...621..445Z, 2009A&A...502..929G, 2011ApJ...740....4Z}, it sets the confidence of positive detections at the level of $99\%$ under the assumption of a Gaussian distribution of the measurements within their associated error bars. Given the median uncertainty of our data ($\pm$0.13\%, Section~\ref{polarimetry}), linear polarization degrees above $\sim$0.39\%~can be detected in both $Z$- and $J$-bands.

Based on this criterion, 7 out of 18 ultracool dwarfs in our sample appear to be linearly polarized in the $J$-band (J0700$+$3157AB, J1159$+$0057, J1411$-$2119, J1501$+$2250, J1521$+$5053, J1807$+$5015, and LP\,415$-$20AB), and 2 out of 5 (J0019$+$4614 and LP\,415$-$20AB) observed in the $Z$-filter appear linearly polarized at 1.03 $\mu$m. This yields a frequency of about 40\,$\pm$\,15\%~of the sample being polarized at 1.03 and 1.25 $\mu$m. This incidence level is similar to that reported by \citetads{2002A&A...396L..35M} and \citetads{2005ApJ...621..445Z} for shorter wavelengths ($I$-band) and dwarfs of similar temperatures and spectral types as our sample.

\begin{figure}
    \includegraphics[width=0.49\textwidth]{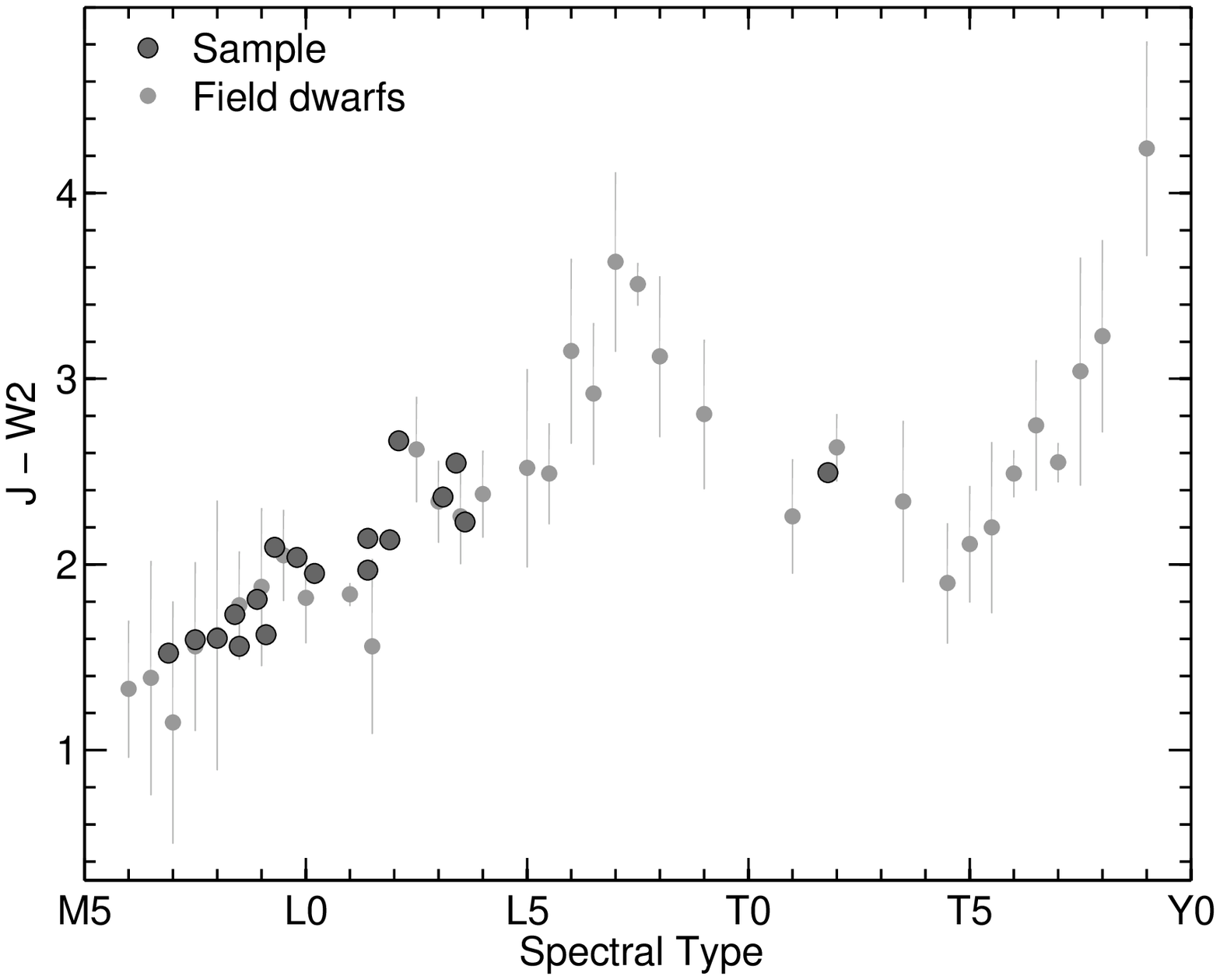}
     \caption{$J-W2$ color as a function of spectral type. Our targets are plotted as black circles; the mean location of field dwarfs is represented by gray circles, where the error bars account for the dispersion of the field \citepads{2012ApJS..201...19D}. Spectral types for our targets are slightly shifted for the clarity of the diagram.}
              \label{Fig3}%
    \end{figure}

Possible scenarios to account for the observed linear polarization are the following:
\begin{itemize}
 \item Interestellar grains in the line of sight towards the targets. \citetads{2002A&A...394..675T} determined that interestellar polarization contributes effectively after 70 pc. As stated in Section~\ref{targets}, our sample is located at closer distances \citeads{2008AJ....136.1290R,2009ApJ...705.1416R,2009AJ....137....1F}, and there is no interstellar extinction reported for any of them. Interstellar polarization is not expected to contribute to our data, and consequently this scenario is rejected. 

 \item Strong magnetic fields can also produce linear polarization by Zeeman splitting of atomic and molecular lines or by synchrotron emission. X-ray and radio observations show that ultracool dwarfs can host magnetic fields in the range 0.1--3\,kG \citepads{1998Sci...282...83N,2005ApJ...627..960B,2006ApJ...648..629B, 2008ApJ...673.1080B, 2010ApJ...709..332B} in agreement with predictions from numerical simulations \citepads{2010A&A...522A..13R}. \citetads{1995A&AS..114...79L} measured typical blue optical ($B$-band) linear polarization degrees of $\lesssim\,0.1\,\%$ in Ap-type stars with magnetic fields of $\sim$1\,kG, finding that the linear polarization amplitude is a function of the intensity of the magnetic field and the number of atomic lines in the stellar spectra. Our targets have spectra fully dominated by the absorption of molecular species, for which the global Zeeman-splitting polarization is expected to be even smaller than for the atomic lines. These low levels of linear polarization are indeed below our detection limit. Persistent radio emission at $\sim$8.5 GHz has been detected in a significant number of cool and ultracool dwarfs \citepads{2012ApJ...746...23M}. In our sample, BRI 0021$-$0214, J0036$+$1821 \citepads{2002ApJ...572..503B}, LP\,349$-$25 \citepads{2012ApJ...746...23M}, and J1501$+$2250 \citepads{2007ApJ...663L..25H} have been detected as radio sources. Assuming gyro- and synchrotron processes associated with their magnetic fields, linear polarization is not expected to be significant at short wavelengths. Of the four radio sources in our sample, only J1501$+$2250 is linearly polarized in the $J$-band. We do not believe that magnetic fields explain our linear polarimetric findings. 

 \item Another source of linear polarization is the presence of dusty material surrounding our targets in the form of proto-planetary disks or debris disks. On the one hand, proto-planetary disks have lifetimes up to $\sim$10\,Myr \citepads{2012ApJ...758...31L} and imprint intrinsic linear polarization degrees of $P\sim1-25\%$  \citepads{2007PASJ...59..487K,2008AJ....136..621K,2009AIPC.1158..111H}. None of our targets have such young ages (Section \ref{targets}), and our polarization measurements are typically below $\sim$0.8$\%$. In Figure \ref{Fig3}, we show the color $J-W2$ as a function of spectral type for our targets and we compare them to the average colors of field dwarfs \citepads{2012ApJS..201...19D}. As seen from the Figure, our objects nicely follow the trend delineated by field dwarfs, indicating no measurable mid-infrared flux excesses. Therefore, there is no evidence for the presence of warm disks around the targets that could account for the observed linear polarization degrees. The scenario of proto-planetary disks is thus rejected for our sample. On the other hand, we cannot discard the presence of debris disks in our sample since these disks are cold and typically detected at wavelengths longer than 4.5 $\mu$m. Those debris disks with particle sizes significantly larger than 1.2 $\mu$m are not expected to produce linear polarization in our filters of interest.

 \item Dust particles present in the upper photospheres or dusty envelopes (e.g., debris disks containing tiny grains of typical size of the order of 1 $\mu$m), causing light scattering processes, are the most likely scenarios to explain the observed linear polarimetric data of the ultracool dwarfs in our sample. Atmospheres have to show inhomogeneities to produce measurable polarization signals as widely discussed by \citetads{2001ApJ...561L.123S} and \citetads{2011ApJ...741...59D}. Inhomogeneities could be related to the presence of nonuniform dusty cloud coverage, and/or to rapid rotation enhancing/generating additional nonuniformity of the atmospheric clouds.
\end{itemize}

%

\subsection{Linear polarization versus rotation}
\begin{figure*}
   \centering
    \includegraphics[width=17cm]{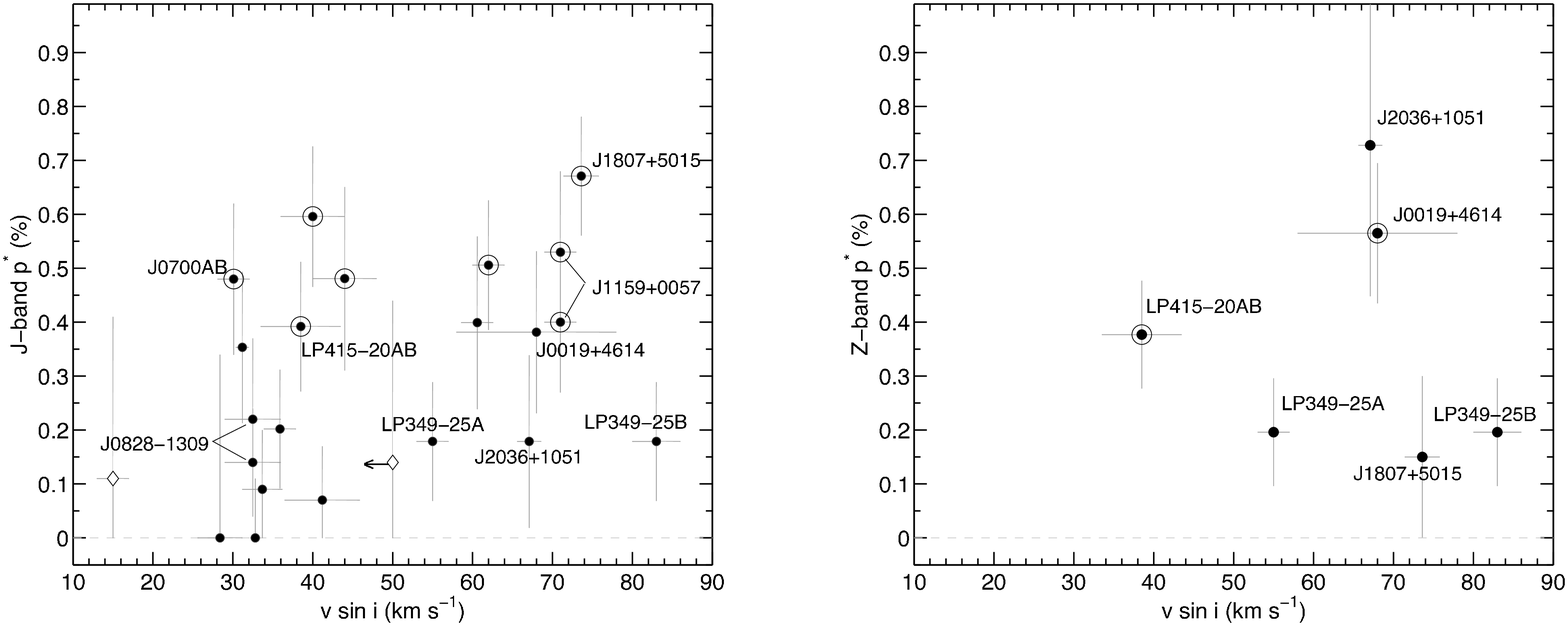}
     \caption{Debiased $J$- ($left$) and $Z$-band ($right$) linear polarization degree as a function of projected rotational velocity. Our data are plotted as black dots and data from the literature (see text) as open diamonds (the arrow stands for an upper limit on $v\,$sin$\,i$). Positive detection of linear polarization ($P/\sigma\ge3$) is indicated by open circles surrounding the black dots. Some objects are labeled. For LP\,415$-$20AB we plotted the average $v$\,sin\,$i$ since both components share similar values; for the LP\,349$-$25 binary system both components are plotted separately given their differing spectroscopic rotational velocities. Velocity measurements and their associated error bars are taken from the literature as explained in the text and Table~\ref{Table1}.}
              \label{Fig4}%
    \end{figure*}

We investigated whether there is a correlation between the observed degree of linear polarization and the projected rotational velocity ($v$\,sin\,$i$). We plot in Figure~\ref{Fig4} the debiased $J$- and $Z$-band polarization degree as a function of $v$\,sin\,$i$ for our sample. Two objects from the literature (J0136$+$0933 and J1022$+$5825, \citeads{2011ApJ...740....4Z}) were added to the $J$-band data because they were observed with the same instrumental configuration as our targets, the uncertainties associated with their polarimetric measurements are $\le \pm0.30$\%, and they have published spectroscopic rotational velocities or photometric rotation periods (see Table 6 by \citeads{2011ApJ...740....4Z}). We obtained their debiased linear polarization degree by applying Equation 2 to the $P$ values given by the authors. Both sources have $v$\,sin\,$i \le 50$ km\,s$^{-1}$ and are unpolarized in the $J$-band.

As shown in Figure~\ref{Fig1} and Table~\ref{Table1}, some of our ultracool dwarfs define the upper envelope of the $v$\,sin\,$i$ versus spetral type diagram. These have $v$\,sin\,$i$ $\ge$ 60\,km\,s$^{-1}$ and rotational velocities approaching the $\sim10-20\%$ of their break-up speed (computed assuming that all ultracool dwarfs have the same size as Jupiter for ages older than $\sim$500 Myr and masses as those given in Section~\ref{targets}). \citetads{2012ApJ...750...79K} reported that some ultracool dwarfs are rotating even at $\sim$ 30\%~of their break-up velocity. The group of 7 targets with $v$\,sin\,$i$ $\ge$ 60\,km\,s$^{-1}$ likely have rotation axis almost perpendicular to the line of sight, i.e., sin\,$i$ $\sim$ 1; the measured spectroscopic rotational velocities reflect their true fast rotation. However, sources with $v$\,sin\,$i$ $<$ 60\,km\,s$^{-1}$ in our sample may present a variety of rotation axis angles, and the uncertainty introduced by the sin\,$i$ prevents us from segregating the very fast rotators ($v \ge 60$ km\,s$^{-1}$) from those rotating moderately ($v$\,=\,30--60 km\,s$^{-1}$).

From the diagram illustrated in Figure~\ref{Fig4}, it is seen that $J$-band linearly polarized sources have $p^{*}$ = 0.4--0.8\,\%~for all rotational velocities. Simple statistics can be performed by including the two objects taken from \citetads{2011ApJ...740....4Z}, and excluding the three binary dwarfs (two of them are polarized) since it is unknown how each component contributes to the measured polarimetric signal. We derived that $50\pm29\%$ (3 out of 6 ultracool dwarfs) in the group of the very fast rotators ($v\,$sin$\,i\ge60$ km\,s$^{-1}$) are linearly polarized in the $J$-band. This contrasts with the frequency of $18\pm13\%$ (2 out of 11) polarized sources among the targets with $v\,$sin$\,i < 60$ km\,s$^{-1}$. The average value of the debiased linear polarization is $<p^{*}>$ = 0.43\,$\pm$\,0.16\,\%~for the very fast rotators, while it is smaller, $<p^{*}>$ = 0.18\,$\pm$\,0.24\,\%, for the dwarfs with moderate spectroscopic rotational velocities. Including the three binaries (each component with its own spectroscopic rotational velocity and the polarimetric measurement of the combined light) does not change the incidence of $J$-band linear polarization significantly: $43\pm25\%$ (3 out of 7) and $29\pm14\%$ (4 out of 14) for the very fast and moderately rotating dwarfs, respectively. We caution that the measured linear polarization degrees of the binaries likely represent a lower limit on the true polarization signal since polarization is ``diluted'' by combining light coming from different sources and conditions. 

Our data suggest that ultracool dwarfs with $v\,$sin$\,i\ge60$ km\,s$^{-1}$ have a higher $J$-band linear polarimetry detection fraction by about a factor of 1.5--2 compared to objects with $v\,$sin$\,i$\,=\,30--60 km\,s$^{-1}$. The number of observations in the $Z$-band is very reduced, and we did not attempt any statistics. However, the $Z$ linear polarimetric data currently available do not appear to contradict the results obtained for the $J$-filter. Nevertheless, the numbers are still small even for the $J$-band observations, and additional data are required for more robust statistics. 

\begin{figure}
   \centering
    \includegraphics[width=0.49\textwidth]{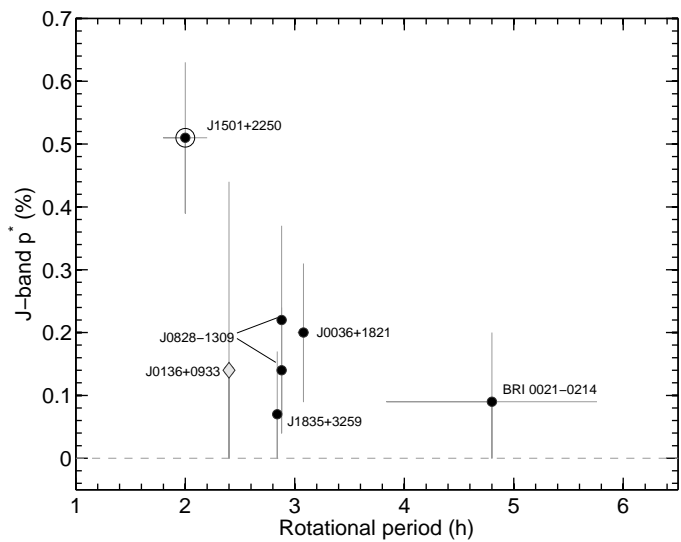}
     \caption{Debiased $J$-band linear polarization degree as a fuction of rotation period. Positive polarimetric detections are indicated with open circles surrounding the black dots. Black dots stand for our measurements, the diamond represents data from the literature (see text). Error bars of rotation periods were computed from the FWHM of the peaks in the periodograms reported in the literature (see Table~\ref{Table1}).}
              \label{Fig5}%
    \end{figure}

Diagrams illustrating $p^{*}$ as a function of true rotational velocity or rotation period are ideal to study the linear polarization dependence on rotation. For five sources in our sample there are rotation periods published in the literature (see Table~\ref{Table1}). And the T2.5 source studied by \citetads{2011ApJ...740....4Z} also has a rotation period determination. These were obtained from the analysis of photometric light curves at wavelengths similar to those of our study. Figure~\ref{Fig5} shows the measured $J$-band linear polarization degree versus rotation period for a total of six ultracool dwarfs with spectral types ranging from M8.5 to T2.5. Note that either true rotational velocity or period can be used without distinction because all ultracool dwarfs in our sample are supposed to have a similar radius. Of the six sources, only one is linearly polarized and it happens to be that with the shortest period or fastest rotation, suggesting that very fast rotation may play a role in the detectability of linear polarization (as suggested by theory). Additional data are demanded to fill in the diagram of Figure~\ref{Fig5} before any solid conclusion can be obtained.

\subsection{Linear polarization versus spectral type}

\begin{figure}
    \includegraphics[width=0.49\textwidth]{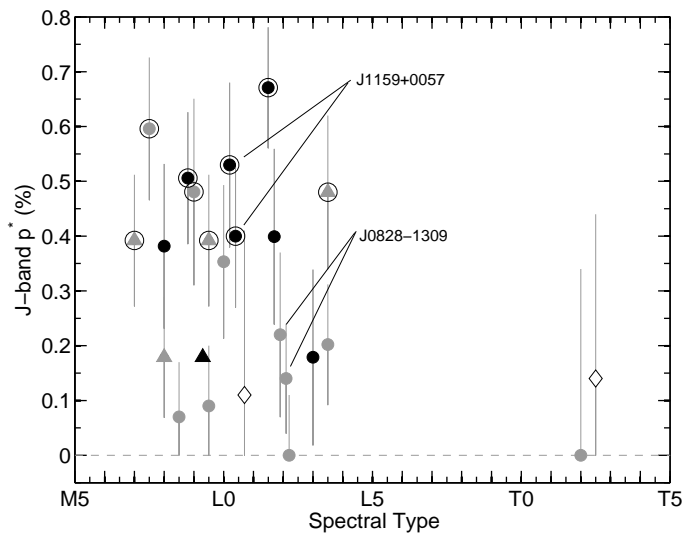}
     \caption{Debiased $J$-band linear polarization degree as a function of spectral type. Objects with $v$\,sin\,$i\ge60$ km\,s$^{-1}$ are plotted as black dots, and sources with slower spectroscopic rotational velocities are indicated with gray dots. Binary objects are indicated with triangles. Encircled symbols stand for linearly polarized sources. The two extra objects taken from \citetads{2011ApJ...740....4Z} ($v$\,sin\,$i<\,$50\,km\,s$^{-1}$) are shown with diamonds.   }
              \label{Fig6}%
    \end{figure}

The debiased linear polarization is shown as a function of spectral type in Figure~\ref{Fig6} using our measurements in the $J$-band. If assumed that the age of the sample is that of the field (typically $\ge$0.5 Gyr, except for J0045$+$1634, discussed below), we may easily relate mass to spectral type in our study: the warmer the spectral classification, the more massive the target object. As indicated in Section~\ref{targets}, the M7--L3.5 sources have likely masses ranging from 0.09 through 0.05 M$_\odot$ and similar surface gravity of log\,$g$ = 5.3--5.1 (cm\,s$^{-2}$). Despite claiming the need for additional data, \citetads{2002A&A...396L..35M} and \citetads{2005ApJ...621..445Z}, argued that there is a slight trend for larger $I$-band linear polarization degree with decreasing surface temperature  for field dwarfs. The spectral type coverage of these previous works expands from the late-Ms to the late-Ls. In our study, no trend is observed for the narrower spectral type interval M7--L3.5; actually, the linear polarimetry scatter appears nearly constant over this spectral range.  

In addition, J1254$-$0122 (T2) is the coolest target in our sample. \citetads{2009ApJ...707..716S} theoretically studied the possible detection of linear polarization in the oblate atmospheres of cloudless T-type sources, which are even cooler than the L-types. Their calculations showed that polarization may arise only for $\lambda\,\le\,0.6\,\mu$m (a range where T--dwarfs are extremely faint), and the net disk integrated polarization may be neglible. We found $p^{*}=0.00\pm0.34\%$ for J1254$-$0122, this measurement and the one obtained for J0136$+$0933 (T2.5; $p^{*}=0.14\pm0.33\%$) by \citetads{2011ApJ...740....4Z} are the only two near--infrared linear polarimetric data available for T-dwarfs; they are in agreement with the theory.

The youngest, lowest gravity (log\,$g$ = 4.2--5.0 cm\,s$^{-2}$ according to the evolutionary models by \citeads{2003A&A...402..701B}), and possibly one of the least massive objects in our sample, J0045$+$1634, is found to be unpolarized ($p^{*}=0.00\pm0.11\%$ in the $J$-band) in contrast with the theoretical predictions of \citepads{2010ApJ...722L.142S}. This may be explained by a true moderate rotation velocity ($v$\,sin\,$i$ = 32.8 km\,s$^{-1}$) or a low inclination angle of the spin axis (i.e., the object is seen near pole-on). To test the linear polarimetric predictions of low gravity (young), cool, low-mass dwarfs made by \citetads{2011MNRAS.417.2874M}, a larger number of observations is demanded.

%
%

\subsection{Linear polarization versus time}
We investigated the variability of the linear polarization degree for objects with measurements available on different occasions. For future references, the Julian date of the observations are provided in Table \ref{Table3}. In the following discussion we separate observations taken at short and intermediate-to-long time scales.

There are three objects (J0019$+$4614, LP\,349--25AB, and LP\,415--20AB) whose $Z$- and $J$-band polarimetric data were taken with a separation of approximately one day. The M8$+$M9 binary LP\,349--25AB has pair members with different rotational velocities, and the contribution of each component to the polarization measurements is not well constrained. The polarimetric observations taken $\sim$1-day apart, or after $\sim$16 rotation cycles, do not reveal significant linear polarization at either 1.03 or 1.25 $\mu$m (to compute the number of rotation cycles we adopted the size of Jupiter, sin\,$i$\,=\,1, and the measured $v$\,sin\,$i$ of the primary given in Table \ref{Table1}. Note that the number of rotation periods estimated in this way always represents a lower limit on the true number of rotation cycles if sin\,$i \ne 1$). The M8 dwarf J0019$+$4614 is linearly polarized in the $Z$-band but does not show polarization at the longest wavelength of our study. This source will be discussed further in the following subsection. The M7$+$M9.5 pair LP\,415--20AB has components with similar spectroscopic rotational velocities, and despite the fact that each member contribution to the polarization is unscontrained, this source shows significant linear polarization degree at both wavelengths. Furthermore, both the intensity of the linear polarization and the vibration angle are nearly constant at the two observing epochs, suggesting that the geometry of the structures responsible for the polarization has likely remained unmodified during at least $\sim$8 rotation cycles.

The ultracool dwarfs J0828$-$1309, J1159$+$0057, J1807$+$5015, and J2036$+$1051 were observed on time scales of months (time intervals of 13.1, 7.5, 3.7, and 3.7 months, respectively). The two later sources were studied at different wavelengths on the two occasions after completing at least 1400 rotation cycles. Because some variability may be expected at long time scales \citepads{2001ApJ...557..822M}, the two observations cannot be compared. On the contrary, J0828$-$1309 (L2) and J1159$+$0057 (L0) were imaged in the $J$-filter after passing a minimum of $\sim$2300 and $\sim$3000 rotation cycles. The former dwarf does not show measurable linear polarization beyond the 3-$\sigma$ detection. The L0 source appears linearly polarized at both observing epochs; however, although the polarimetric degree seems unchanged within the quoted uncertainties, the polarization vibration angles look quite different. This can be interpreted as follows: the amounts of photospheric dust/condensates responsible for the light scattering processes may be similar at the two epochs, but the the location or distribution of the heterogeneous ``cloud'' coverage may have evolved across the observable disk.

Based on LP\,415--20AB and J1159$+$0057, our data hint at small linear polarimetric variability amplitudes on short time scales, i.e., about days or a few ten rotation cycles, and larger polarimetric varibility on longer time scales, i.e., over hundred to thousand rotation periods. Theory predicts that low-mass objects with $M\le0.35$ M$_{\sun}$ become fully convective \citepads{2000ARA&A..38..337C}. The very fast rotation has an impact in the dwarfs convection processes \citepads{2012arXiv1210.7573S}; the models produced by these authors suggest that, when cloud condensation levels lie in the atmosphere (e.g., late-M and L dwarfs), patchy clouds will form leading to variability that is not canceled out in a disk integrated light. According to \citepads{2012arXiv1210.7573S}, for a typical brown dwarf with a rotation period of $\sim$10$^{4}$ s, changes in the detailed structure of the atmospheric turbulence occur on time scales of about $\sim$10--100 rotation cycles. This is qualitatively in agreement with our (yet scarce) polarimetric observations, and with the photometric light curves of some late-M and L dwarfs available in the literature. For example, \citetads{2001ApJ...557..822M,2009ApJ...701.1534A} and \citetads{2012ApJ...750..105R}  reported on ultracool dwarfs photometric variations not only on rotational time scales (a few hours) but also over many rotation cycles. The polarimetric monitoring along with simultaneous photometric light curves may provide new insights into the ``weather'' of ultracool dwarfs.

\begin{table}
\caption{Other polarimetric measurements.}
\label{Table5}
\centering
\begin{tabular}{l c c c c}
\hline\hline
Object&$\lambda_{c},\delta\lambda$$^{\rm a}$&$p^{*}$&$\Theta$&Ref.\\
  &($\mu$m)&($\%$)&(deg)& \\
\hline
BRI0021$-$0214 &0.85,0.15 &0.06$\pm$0.17 &--&2\\
J0036$+$1821   &0.64,0.16 &0.52$\pm$0.33 &--&2\\
               &0.77,0.14 &0.20$\pm$0.03 &17.6$\pm$4.0&1\\
               &0.77,0.14 &0.03$\pm$0.05 &--&3\\
               &0.85,0.15 &0.00$\pm$0.06 &--&2\\
J0045$+$1634   &0.85,0.15 &0.00$\pm$0.12 &--&2\\
J1807$+$5015   &0.85,0.15 &0.00$\pm$0.06 &--&2\\
               &0.81,0.14 &0.70$\pm$0.14 &46.7$\pm$1.2&4\\
               &0.64,0.14 &1.55$\pm$0.61 &101$\pm$1.2&4\\
J1835$+$3259   &0.85,0.15 &0.04$\pm$0.03 &--&2\\
J2057$-$0252   &0.77,0.14 &0.04$\pm$0.02 &--&1\\
               &0.85,0.15 &0.00$\pm$0.38 &--&2\\
\hline
\end{tabular}
\tablefoot{
\tablefoottext{a}{Passband central wavelength and width as reported by the authors.}
}
\tablebib{
(1) \citetads{2002A&A...396L..35M}; (2) \citetads{2005ApJ...621..445Z}; (3) \citetads{2009A&A...502..929G}; (4) \citetads{2009A&A...508.1423T}.
}
\end{table}

Six of our targets have published linear polarimetric data at optical wavelengths ($R$- and $I$-bands). In Table~\ref{Table5} we compiled all available measurements by providing the central wavelengths and widths of the passbands of the observations, the debiased linear polarization degrees, the polarization vibration angles (in case of positive polarimetric detections), and the bibliographic references. These data were acquired years before our observations and were taken using different filters, instruments, and telescopes. BRI\,0021$-$0214, J0045$+$1634, J1835$+$3259, and J2057$-$0252 do not show linear polarization at either optical or near-infrared wavelengths on different epochs of observations. The L3.5 dwarf J0036$+$1831 was reported to be linearly polarized in the Bessel $I$-band by \citetads{2002A&A...396L..35M}, while 4 yr later \citetads{2009A&A...502..929G} did not detect polarization above the 3-$\sigma$ level in observations taken with the same instrumental configuration as the first data. Images collected at longer wavelengths in the present work and in \citetads{2005ApJ...621..445Z} did not reveal any significant linear polarization. The L1.5 dwarf J1807$+$5015 is reported to be polarized and unpolarized at different wavelengths and epochs, suggesting strong polarimetric variability. This object is further discussed in the next subsection.

%
   \begin{figure}
    \includegraphics[width=0.49\textwidth]{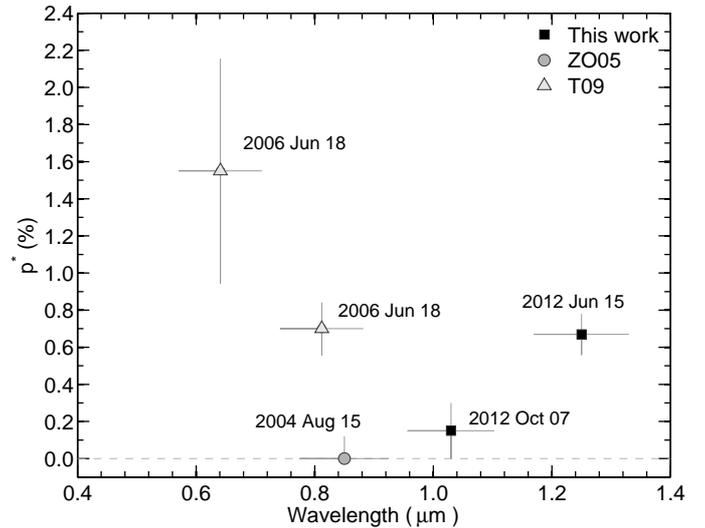}
     \caption{Debiased linear polarization degree  measured for J1807$+$5015 as a function of wavelength. Data were compiled from \citetads{2005ApJ...621..445Z}, \citetads{2009A&A...508.1423T}, and this work. Vertical error bars correspond to the quoted uncertainties in polarization, and the horizontal error bars account for the width of the filters. }
              \label{Fig7}%
    \end{figure}

\subsection{Linear polarization versus wavelength}

The degree of linear polarization produced by light scattering increases significantly when the size of the grain particles is comparable to the wavelength of the observations \citepads{2001ApJ...561L.123S}. Therefore, a multiwavelength study of ultracool dwarfs linear polarization may provide relevant information on the typical sizes of the atmospheric grains/condensates, which is a key ingredient for the theory of model atmospheres (e.g., \citeads{2001ApJ...556..357A}). We caution that the linear polarization degree is also a function of the wavelength-dependent gas opacity: in atmospheric dusty regions with strong gas opacity there will be less scattering processes and less polarization (see discussion in \citeads{2011MNRAS.417.2874M}). Consequently, linear polarization measurements reflect a combination of various factors: gas opacity function, presence of dust, and grain physical sizes. In addition, if polarimetric observations are not simultaneous, variability of dusty patterns may become an issue. Our sample objects are brighter in the $J$-band than at shorter wavelengths, i.e., gas opacity is reduced at around 1.2 $\mu$m. Unfortunately, none of our $Z$- and $J$-band observations are simultaneous.

By combining our observations and the data from the literature we may draw some preliminary results. The L1.5 source J1807$+$5015 has the largest number of polarimetric measurements among all studied ultracool dwarfs. Figure \ref{Fig7} illustrates the debiased linear polarization degree as a function of wavelength from the $R$- through the $J$-bands (Tables~\ref{Table3} and~\ref{Table5}). Despite the fact that gas opacity is stronger at the $R$-band wavelengths, linear polarization intensity is found to be the largest at the shortest wavelength, likely implying submicron atmospheric particles as shown in Figure~7 of \citetads{2005ApJ...625..996S}. The two measurements taken in the $I$-band (although slightly different filters, Table~\ref{Table5}) differ by $\delta p^{\ast}$\,$\sim$0.7\,\%~and were obtained on two occasions separated by nearly 2 yr. We ascribe this polarimetric difference to a likely atmospheric variability. The high value of $p^{\ast}$ obtained for the $J$-band as compared to the $Z$-filter in our work cannot be attributed only to differing gas opacities since more than $\sim$1400 rotation periods were completed between these two observations and changes are expected to occur in atmospheric dusty structures \citepads{2010A&A...513A..19F}. 

The predominant submicron-to-micron size of the dusty particles is also supported by the observations of J0019$+$4614 (M8) and J2036$+$1051 (L3), whose linear polarization indices at 1.03 $\mu$m are larger than those at 1.25 $\mu$m, and the observations of LP\,415$-$20AB, which show similar polarization degrees at both wavelengths.  Simultaneous multiwavelength polarimetric observations may help constrain model atmospheres since the models available in the literature tend to agree on the global atmospheric structure but differ in details like grain size among other opacity-related parameters (e.g., see \citeads{2008MNRAS.391.1854H}).

\section{Conclusions \label{conclusions}}
Using the LIRIS instrument on the William Herschel telescope, we obtained $Z$- (1.03 $\mu$m) and $J$-band (1.25 $\mu$m) linear polarimetric images of a sample of 18 bright, rapidly rotating ultracool dwarfs with $v$\,sin\,$i\gtrsim$\,30 km$\,$s$^{-1}$ and spectral types ranging from M7 through T2. All these sources are believed to form condensates of liquid and solid particles in their atmospheres and to have rather oblate shapes given their large rotational velocities (some are rotating at nearly one third of their break-up velocity). This provides an appropriate scenario for the detection of linear polarization in the continuum light produced by scattering processes.  Three of the targets are known binaries. The median uncertainty associated with our $Z$- and $J$-band measurements was $\pm$0.13\,\%~in the linear polarization degree, indicating that our data are sensitive to strong polarization indices $p^{\ast}$\,$\ge$\,0.39\,\%~at the $\ge$3-$\sigma$ confidence level. 

Eight ultracool dwarfs out of 18 targets appear to be linearly polarized in the $J$- and/or $Z$-bands, suggesting that about 40\,$\pm$\,15\,\%~of the sample shows significant linear polarization in the near-infrared. Measured positive detections have linear polarization degrees in the range $P$\,=\,0.4--0.8\,\%~independently of spectral type and $v$\,sin\,$i$. Our derived polarimetric degrees are in agreement with theoretical predictions \citepads{2010ApJ...722L.142S,2011ApJ...741...59D}. Additionally, our data hint at ultracool dwarfs with the highest rotation ($v\,$sin$\,i\ge60$ km\,s$^{-1}$) having a factor of about 1.5--2 larger $J$-band linear polarimetry detection fraction than ultracool dwarfs with $v\,$sin$\,i$\,=\,30--60 km\,s$^{-1}$. The average value of the debiased linear polarization (including detections and non-detections) is $<p^{*}>$ = 0.43\,$\pm$\,0.16\,\%~for the very fast rotators, while it is smaller, $<p^{*}>$ = 0.18\,$\pm$\,0.24\,\%, for the dwarfs with moderate spectroscopic rotational velocities. There are  six objects with true rotation period in the literature, the one with the shortest period (2 h) and the largest $v$\,sin\,$i$ (60 km\,s$^{-1}$) shows linear polarization in the $J$-band. Further data are required for a more robust statistics.

We also investigated the dependence of $J$-band linear polarimetry on time and wavelength for the ultracool dwarfs in our sample. For those objects with polarimetric observations obtained $\sim$1-day apart (or after completion of about a few ten rotation periods), our data suggest little polarimetric varibility (both in the linear polarization degree and vibration angle). Observations taken months apart (over hundreds to thousand rotation cycles) display significant polarimetric variability, suggesting changes in the atmospheric structures responsible for the observed linear polarization. For those sources with polarimetric detections in $Z$- and $J$-bands on short time scales, we observed that the linear polarization degree tends to be larger at the shortest wavelength, implying sub-micron (or around the micron) grain sizes. Simultaneous observations of photometric and polarimetric light curves of variable sources will shed new light in our comprehension of the ``weather'' affecting ultracool dwarfs and the sizes of the atmospheric dusty particles.

\begin{acknowledgements}
We are thankful to the referee, Prof. Mark Marley, for his valuable report. The William Herschel Telescope is operated on the island of La Palma by the Isaac Newton Group in the Spanish Observatorio del Roque de los Muchachos of the Instituto de Astrof\'\i sica de Canarias. This research has made use of NASA's Astrophysics Data System and the SIMBAD database, this last one being operated at CDS, Strasbourg, France. Also, this publication makes use of data products from the Wide-field Infrared Survey Explorer, which is a joint project of the University of California, Los Angeles, and the Jet Propulsion Laboratory/California Institute of Technology, funded by the National Aeronautics and Space Administration. This work is partly financed by the Spanish Ministry of Economics and Competitiveness through projects AYA2010-21308-C03-02 and AYA2011-30147-C03-03.
\end{acknowledgements}
\bibliographystyle{aa} 
\bibliography{biblio} 
\end{document}